# Transmit Antenna Subset Selection In Mimo Ofdm System Using Adaptive Mutuation genetic Algorithm

Nidhi Sindhwani[1] and Manjit Singh[2]

[1]ASET, GGSIP University, New Delhi, India
[2]UCOE, Punjabi University, Patiala, India

## Abstract

*Multiple input multiple output techniques are considered attractive for future wireless communication systems, due to the continuing demand for high data rates, spectral efficiency, suppress interference ability and robustness of transmission. MIMO-OFDM is very helpful to transmit high data rate in wireless transmission and provides good maximum system capacity by getting the advantages of both MIMO and OFDM. The main problem in this system is that increase in number of transmit and receive antennas lead to hardware complexity. To tackle this issue, an effective optimal transmit antenna subset selection method is proposed in paper with the aid of Adaptive Mutation Genetic Algorithm (AGA). Here, the selection of transmit antenna subsets are done by the adaptive mutation of Genetic Algorithm in MIMO-OFDM system. For all the mutation points, the fitness function are evaluated and from that value, best fitness based mutation points are chosen. After the selection of best mutation points, the mutation process is carried out, accordingly. The implementation of proposed work is done in the working platform MATLAB and the performance are evaluated with various selection of transmit antenna subsets. Moreover, the comparison results between the existing GA with mutation and the proposed GA with adaptive mutation are discussed. Hence, using the proposed work, the selection of transmit antenna with the maximum capacity is made and which leads to the reduced hardware complexity and undisturbed data rate in the MIMO-OFDM system*

## 1. Introduction

MIMO techniques have quite large potential for future wireless communication systems, due to the ever increasing demand for high data rates and spectral efficiency [7]. There has been substantial benefit in multiple-input multiple-output (MIMO) systems that utilize more than one antenna at each terminal. The channel capacity can sufficiently be improved through the additional degrees of freedom available with multiple antenna systems. Theoretically, the capacity of MIMO systems is directly proportional to the minimum number of transmit and receive antennas [12].OFDM is a spectrally effective modulation that converts a frequency selective channel into a set of parallel frequency-flat channels and allows simple equalization schemes if the channel length is less than the length of the cyclic prefix. Space-time signaling with numerous antennas per transceiver has been shown to a remarkable increase in capacity and provide considerable diversity especially when there is channel knowledge at the transmitter [16-17]. OFDM approach is expected to improve performance in combating disadvantages in frequency selective fading encountered in MIMO wireless systems [7].The substantial drawback in MIMO-based systems is the hardware cost, because every antenna element need a complete radio frequency (RF) chain which contains mixers, amplifiers, and analog-to-digital conversion. Antenna subset selection is a potential technique has been proposed to simplify the hardware complexity, i.e., save on RF chains, while providing many heterogeneous advantages [8-9]. An antenna-subset-selection-based MIMO system utilizes a number of RF chains, each of which is switched to operate numerous antennas. The throughput/reliability tradeoff can also be increased by antenna selection techniques along with reducing the system cost [13].In antenna subset selections, the number of RF chains is less than the actual number of antenna elements. The RF chains are connected to the "optimum" antenna elements, where "optimum" depends on the channel state (i.e., can vary with time).

## 2. MIMO OFDM SYSTEM

A general structure of MIMO-OFDM system contains two kinds of antennas, transmit antennas and receive antennas are represented as $N_t$ and $N_r$ . Moreover, frequency selective channel is utilized in this system. The structure of this MIMO-OFDM system is given in the following figure 1. In this system it is assumed that the number of sub-carriers be $N_s$. For every subcarrier $i$ , consider the signals such as transmitted signal and received signal are $\iota_i$ and $r_i$ and also the Additive White Gaussian Noise (AWGN) is $g_i$. If the sub-carrier $i$ of size $N_t \times N_r$ has the channel response matrix be $C_i$, then the received signal can be represented as follows in eqn. (1).

$$r_i \equiv g_i + s_i \qquad\qquad (1)$$

In this equation (1), the realization of the Gaussian random matrix is known at the side of receiver. And this is represented as in equation (2)

$$C_i \equiv \sum_{j=0}^{L-1} c_j e^{-j2\pi i / N_s} \qquad\qquad (2)$$

Where, $c_i$ an uncorrelated channel matrix, in which every matrix elements follows IID (Independently and Identically Distributed) complex Gaussian distribution. $L$- L-tap frequency selective channel, which indicates the tap of selected channel. The Channel State Information (CSI) is available on the receiver side only and not on the transmitter side. For all the space-frequency sub-channels, whole available power in the system is allocated equally. Between various time slots, we can work out with the system by selecting the antennas. For a particular time slot, we can work with only one subset of $N_t$ transmit antennas. One of the feedback channels named as error free and delay, which helps to send the antenna subset index from the receiver to the transmitter. From the values of identity matrix (for both transmitter and receiver), SNR and the number of transmit antennas that are utilized in a subset, we can find the ergodic capacity level for a particular antenna to be selected.  Let $I_{N_t}$ and $I_{N_r}$ be the identity matrix of size $N_r \times N_t$ and $N_r \times N_r$ respectively, $\rho$ be the SNR value for every sub carrier and $N_t$ be the number of transmit antennas used in a subset. Then the following equation gives the ergodic capacity for the antenna selection process.

$$EC = E\left[\frac{1}{Ns}\left[\log\left[\sqrt{I_{N_r} + \left(\frac{\rho}{n_t}\right)(I_{N_t})(H_0 H_0^H)} + \sqrt{I_{N_r} + \left(\frac{\rho}{n_t}\right)(I_{N_t})(H_1 H_1^H)} + \cdots \sqrt{I_{N_r} + \left(\frac{\rho}{n_t}\right)(I_{N_t})(H_{N_s-1} H_{N_s-1}^H)}\right]\right]\right] \qquad (3)$$

In equation (3), $H_i^H$ denotes the conjugate transpose of $H_i$.

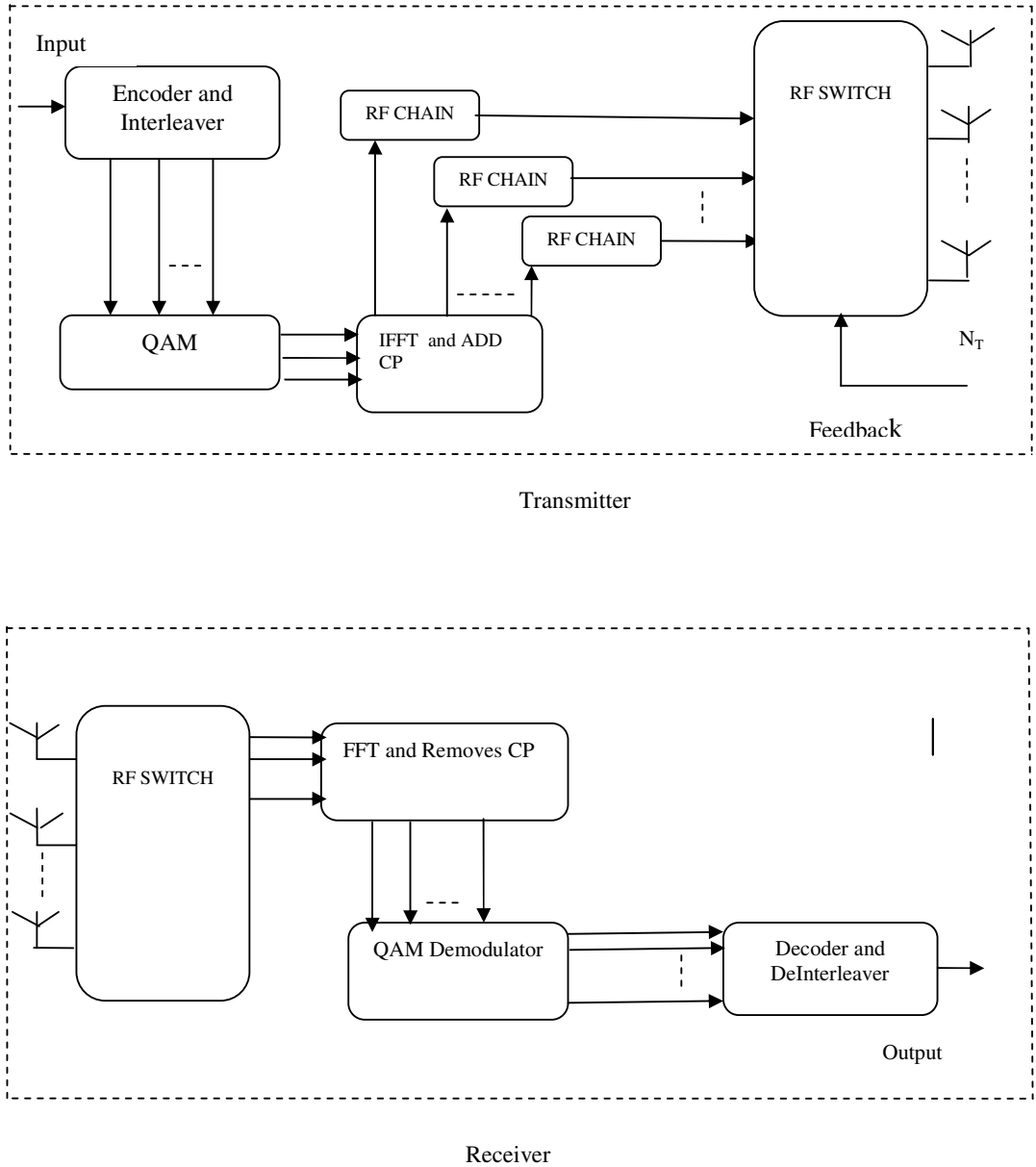

Figure 1.  General Structure of MIMO-OFDM system (a) Transmitter (b) Receiver

# 3. ADAPTIVE GENETIC ALGORITHM

Several antenna selection algorithms and techniques have been proposed for solving the mentioned dilemma in the MIMO-OFDM system so far, yet the results of those algorithms were not satisfactory. More number of transmit antennas and receive antennas can increase the complexity of hardware in this system. This limitation is overcome by the proposed work to bring out better channel capacity with decreasing in hardware complexity. The proposed technique utilizes Genetic algorithm with adaptive mutation. As the mutation, one of genetic operations is made adaptive; the solution can be converged with less time rather than the Genetic algorithm with conventional mutation. As mutation plays a key role in all the genetic operators, introducing adaptive techniques in mutation overcomes the difficulties in identifying the optimal solution. They not only triumph over the difficulties in finding the optimal solutions, but also adapt the mutation rate well according to the evolutionary process. Also the solution obtained is not confined to any limited problem and so it makes the algorithm more flexible to handle any of such antenna selection problems. Finally, optimal antennas are selected using the proposed transmit antenna selection so that a good system capacity is achieved with less hardware complexity. The Proposed method for the Genetic Algorithm with adaptive mutation is shown in figure 2.

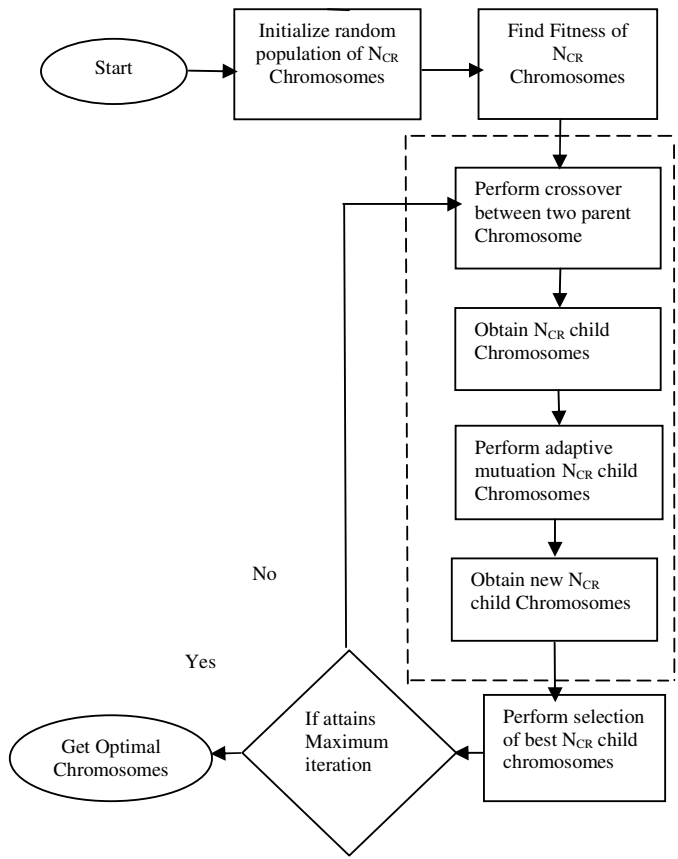

Figure 2. Proposed method diagram for the Genetic Algorithm with Adaptive Mutation

## 3.1. Genetic algorithm with adaptive Mutation for the selection of transmit antennas

### *Initialization*

At first, $N_{CR}$ numbers of chromosomes are generated randomly in GA with the length of $N_t$. The random generated chromosomes are represented as follows

$$CR^{(k)} = \left\{ cr_0^{(k)} \; cr_1^{(k)} cr_2^{(k)} \dots \dots \dots cr_{Ni-1}^{(k)} \right\} \qquad o \le k \le N_{CR} - 1 \qquad (4)$$

In above equation (4), $N_t$ numbers of genes are presented in a chromosome and every gene has a value '0' or '1' with it as given in following equation (5).

$$cr_n^{(k)} = \{ 1 \; if \; n^{(k)} \subset [P^{(k)}] \qquad\qquad (5)$$

Otherwise it will take as 0. Where, $cr_n^{(k)}$ - $n_{th}$ gene of the $k_{th}$ chromosome and the number of genes are $N_t$. $[P^{(k)}]$ — a vector that indicates the position of every gene.

Based on random integer for chromosome, the position of the genes are gained as,

$$[P^{(k)}] << RI_a^{(k)} \% N_t \qquad\qquad (6)$$

From equation. (6), we can understand that a gene's position is found from $a_{th}$ random integer for $k_{th}$ chromosome $RI_a^{(k)}$

### *Fitness Calculation*

After that, the $N_{CR}$ chromosome values are computed by fitness value as in equation. 7.

$$F_k = arg \max_{cr^{(k)}} EC\left( CR^{(k)} \right) \qquad\qquad (7)$$

In this equation (7), the ergodic capacity value $EC(CR^{(k)})$ is applied for the $k_{th}$ chromosome and the value of $EC(CR^{(k)})$ is specified given below in equation (8) as in the equation (3).After finding the fitness values for all the $k$ chromosomes, the next processes of genetic algorithm are subsequently continued.

$$EC(CR^{(k)}) = E\left[ \frac{1}{NS} \log \left| \sqrt{I_{N_r} + \left(\frac{\rho}{n_t}\right)(I_{N_t})\left(H_0^{(k)}\left(H_0^{(k)}\right)\right)^H} + \sqrt{I_{N_r} + \left(\frac{\rho}{n_t}\right)(I_{N_t})\left(H_1^{(k)}\left(H_1^{(k)}\right)\right)^H} + \dots \sqrt{I_{N_r} + \left(\frac{\rho}{n_t}\right)(I_{N_t})\left(H_{N_t-1}^{(k)}\left(H_{N_t-1}^{(k)}\right)\right)^H} \right| \right] \qquad (8)$$

### *Crossover*

The single point crossover operation is carried out in this crossover operation with the crossover rate of 0.5. The parent chromosomes subjected to crossover with them and the children chromosomes are attained from the result of the process. The genes that are placed right to the crossover point in the parent chromosomes are interchanged between the parents The interchanged genes between the parent chromosomes produce the children chromosomes.Thus, $N_{CR}$ numbers of children chromosomes are obtained from $N_{CR}$ numbers of parent chromosomes using crossover process. The representation of children chromosomes is

$$CR_{child}^{(k)} = \left\{ cr_0^{(k)} \; cr_1^{(k)} cr_2^{(k)} \dots \dots \dots cr_{Ni-1}^{(k)} \right\} \qquad N_{CR} \; o \le k \le 2N_{CR} - 1 \qquad (9)$$

Thus the children chromosomes are obtained from the crossover operation and which are given as the input to the Adaptive mutation process of Genetic Algorithm.

### 3.1.1. Adaptive mutation for subset transmit antenna selection

In this work, the selection of subset transmit antenna is made by the adaptive work of mutation process in the Genetic algorithm. For a particular mutation point, the process is continuously run for more times to make the best chromosomes. This is based on the fitness value of mutation point of every child chromosomes. This process is not only for one mutation point, but for all the mutation points and then the fitness value based best point is selected. The diagram for adaptive mutation is given in figure 3.Each mutation points are found by their child chromosomes and the feasible mutation points are indicated by 1s and others by 0s. The number of feasible mutation points are $N_{MU}^{(l)}$ that indicates the number of mutation points (number of 1s) in a chromosome. The feasible mutation point representation is given below in equation (10).

$$MU_f^{(l)} = \left\{ mu_{f0}^{(l)} mu_{f1}^{(l)} mu_{f2}^{(l)} \ldots \ldots mu_{fN_i-1}^{(l)} \right\} 0 \le f \le N_{MU}^{(l)} - 1 \qquad (10)$$

Based on the position of genes $P_{MU}^{(l)}$ with 1s, each genes in a feasible mutation point are known with the following (eqn. 11) condition. Moreover, the position of genes are placed between the ranges $\left( 0 \le P_{MU}^{(l)}(f) \le N_t - 1 \right)$ and $P_{MU}^{(l)} = N_{MU}^{(l)}$

$$. MU_{fn}^{(l)} = \left\{ 1; if \ n^{(l)} \in \left[ P_{MU}^{(l)} \right] \right. \qquad (11)$$

Else it will take 0. Thus, feasible  mutation points for every child chromosomes is obtained. Apply the following fitness evaluation (equation (12)) as in equation (7) to all these feasible mutation points for identifying the best subset transmit antennas. Based on the ergodic capacity value, the fitness function is calculated as follows

$$F_{MU}^{(l)}(f) = arg \max_{MU_f^{(l)}} EC \left( MU_f^{(l)} \right) \qquad (12)$$

Then the best values of fitness are chosen as the best mutation points for selecting the optimal subset transmit antennas.

$$MU_{best}^{(l)} = Best_{M_f^{(l)}} \left( F_{MU}^{(l)}(f) \right) \qquad (13)$$

Thus after applying the fitness value for each mutation points, the best one mutation point can be obtained and the mutation process over the best mutation points again generates new child chromosomes. These new child chromosomes are also subjected to the fitness function for its evaluation of best point

### _Selection_

A selection pool is made with $N_{CR}$ parent chromosomes and  $N_{CR}$ child chromosomes, according their fitness value. The order for arranging the chromosomes is also based on the fitness value and the best fitness providing chromosomes are placed in topmost level. For the process of next generation, the first $N_{CR}$ chromosomes in the topmost level of selection pool are chosen over the collection of $N_{CR}$ parent chromosomes and $N_{CR}$ child chromosomes. Using the chosen chromosomes, the genetic algorithm process starts from the operator crossover and finishes when the termination criterion reaches. The whole GA process gets finished, when the maximum generation of chromosomes occurred and the best chromosomes are chosen from the topmost level of selection pool.

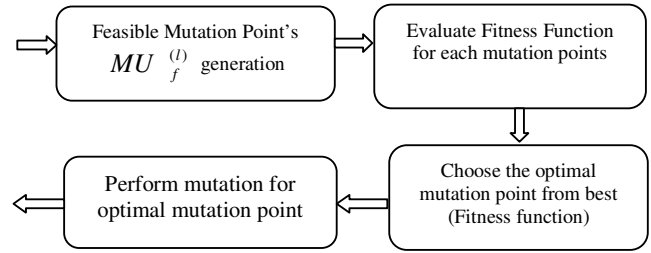

Figure 3. Adaptive mutation steps in Genetic Algorithm

## 4. RESULT AND DISCUSSIONS

In this proposed GA with adaptive mutation is run for a particular SNR to select the optimal $n_t$ subset transmit antennas. Hence, with the aid of the proposed technique for transmit antenna subset selection; a good system capacity with reduced hardware complexity with good ergodic capacity and undisturbed data rate is accomplished in the MIMO-OFDM system. Proposed work for the selection of subset transmit antenna in MIMO-OFDM system using Genetic Algorithm with adaptive mutation is implemented in the Matlab (version 8.0.0.783 (R2012b)) platform.

### 4.1. Evaluation of MIMO-OFDM system

The evaluation of MIMO-OFDM system with the proposed work starts by considering the transmit antennas and receive antennas. The number of transmit antennas used are ten and the number of receive antennas used for work is also ten. $i.e. N_t = 10 \ and \ N_r = 10.$ The aim of proposed work is to choose best subset of transmit antennas. By changing the number of subsets, proposed work is evaluated with the ergodic capacity values of every number of subsets. The varying numbers of subsets of optimal transmit antennas are $i.e. n_t = 2, 4, 6, 8$. For a given SNR value ($\rho$), the optimal transmit antenna subsets are selected and the ergodic capacity for every selected subsets are evaluated based on it, which is tabulated in the table 1

As it is identified that when the number of selected optimal transmit antennas increases, the value of ergodic capacity for the corresponding antennas get increases. In this work, from 10 transmit antennas, the number of selected optimal transmit antennas are varied as 2, 4, 6 and 8. For $n_t = 2$, the selected optimal antennas are 4 and 10; for $n_t = 4$, the selected optimal antennas are 1, 2, 5 and 7; for $n_t = 6$, the selected optimal antennas are 1, 2, 3, 6, 8 and 9; And finally for, $n_t = 8$ the selected optimal antennas are 2, 3, 4, 6, 7, 8, 9 and 10. Moreover, the ergodic capacity values are also found from the proposed work at the value of $\rho = 15 \ dB$. The ergodic capacity values for $n_t = 2, 4, 6 \ and \ 8$ are 17.6108 bits/s/Hz, 28.9241 bits/s/Hz, 39.5586 bits/s/Hz and 47.4062 bits/s/Hz, respectively, which indicates that the value of ergodic capacity is increased for the higher number of $n_t$. Based on the number of $n_t$, the changes in ergodic capacity is happened. the capacity is highly increased with the usage of the proposed work and it leads to reduce the hardware complexity.

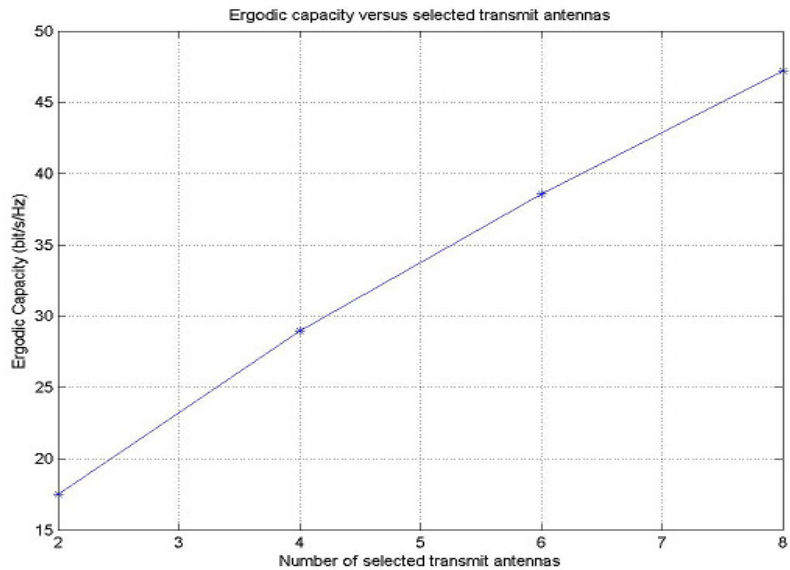

Figure. 4 Ergodic Capacity versus SNR for different number of transmit antenna i.e. . $n_t = 2, 4, 6, 8$

The Ergodic capacity values for various SNR values with the number of selected optimal transmit antennas at the value of $l = 3$ is plotted in the following graph in figure 5. Here, the values are plotted for various antenna subsets ($n_t = 2, 4, 6, 8$). When the optimal transmit antenna subsets are in small number ($n_s = 2$), the ergodic capacity is very low and when the optimal transmit antenna subsets are in big number ($n_t = 8$), high value of ergodic capacity is obtained. And also, if the SNR value is small means, very good capacity value cannot be gained and when the SNR value is high, higher ergodic capacity values can be attained

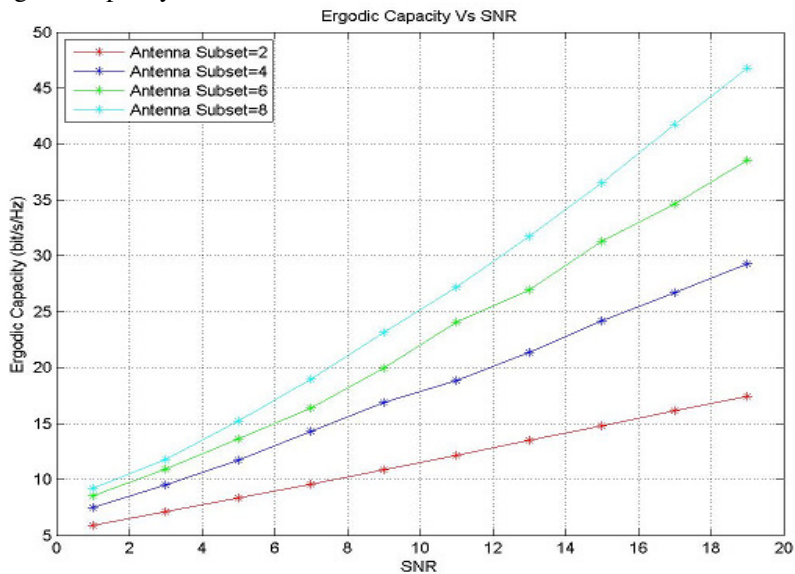

Figure 5. Ergodic Capacity versus SNR for different number of transmit antenna
for $N_t = 10, N_r = 10 \ and \ \rho = 15dB$

The performance results of optimal transmit antenna selection using the proposed GA-adaptive mutation is evaluated with the 10 total number of transmit antennas and 8 array subsets of transmit antennas at the rate $\rho = 15\ dB$. From the graph values in figure 6, it can be identified that the performance of GA-adaptive mutation results. At the first iteration, the value of ergodic capacity is 0 bit/s/Hz. After that, from the second iteration onwards, the ergodic capacity value moves to the higher values and finally it reaches the value of 47.4062 bit/s/Hz. From these results, it is easily observable that the ergodic capacity value is very highest for all iterations except the 1[st] iteration in GA-adaptive mutation and thus this proposed work facilitates good ergodic capacity results with the best transmit antennas subset selection for the reduction in hardware complexity by providing the adaptive mutation with best fitness evaluation value for all mutation points.

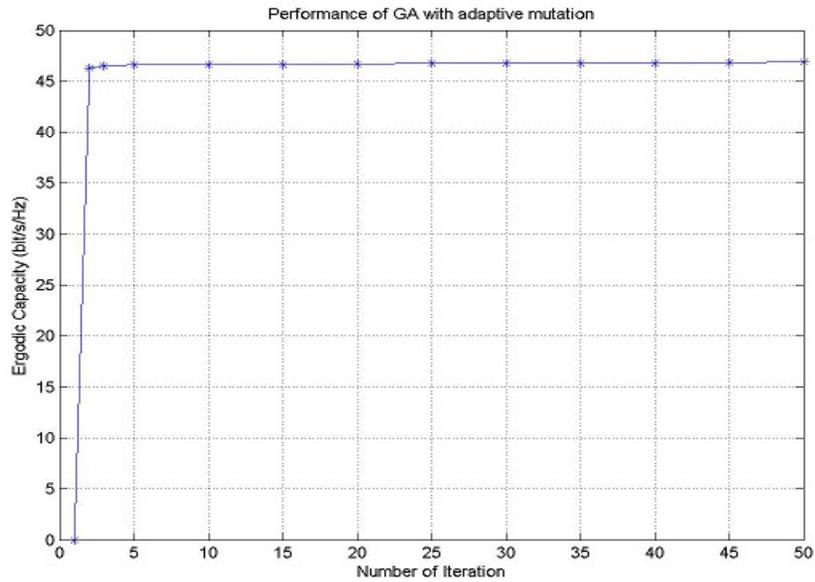

Figure 6. Performance of GA while using adaptive mutation responds to convergence with
$N_r = 10, N_r = 10 \ and \ \rho = 15dB$

## 4.2. Comparison of proposed GA with adaptive Mutation work with existing GA

From Section 4.1, it is clearly notified that the proposed GA-adaptive mutation work can have the ability to make good ergodic capacity with the optimal transmit antennas subset selection.It is not enough to prove that the proposed work is best one.In order to prove with more results, there is a comparison between proposed approach with the existing GA-mutation work in MIMO-OFDM. The table 2 values give the results of comparison.

Table-2: Comparison table for Optimal antenna subset selection and resultant ergodic capacity of the system for $N_t = 10, N_r = 10 \ and \ \rho = 15dB$

| Number of Subcarriers ($n_t$) | Number of transmit Antenna ($N_t$) | Selected optimal antennas | | Ergodic capacity (bits/s/Hz) | |
|---|---|---|---|---|---|
| | | GA-adaptive mutation | GA-mutation | GA-adaptive mutation | GA-mutation |
| 2 | 10 | 4,10 | 5,7 | 17.6108 | 14.5045 |
| 4 | 10 | 1,2,5,7 | 1,3,4,6 | 28.9241 | 23.9472 |
| 6 | 10 | 1,2,3,6,8,9 | 3,4,5,8,9,10 | 39.5586 | 32.0696 |
| 8 | 10 | 2,3,4,6,7,8,9,10 | 1,3,4,5,7,8,9,10 | 47.4062 | 37.4518 |

The subsequent graph in figure7 illustrates the comparison graph of GA-mutation & GA-adaptive mutation for the results of Ergodic capacity versus number of selected optimal transmit antennas The performance results of optimal transmit antenna selection using proposed GA-adaptive mutation is evaluated with the 10 total number of transmit antennas and 8 array subsets of transmit antennas at the rate $\rho = 15dB$. From the graph values in figure 7, it can be identified that the performance of GA-adaptive mutation results. At the first iteration, the value of ergodic capacity is 0 bit/s/Hz. After that, from the second iteration onwards, the ergodic capacity value moves to the higher values and finally it reaches the value of 47.4062 bit/s/Hz. From these results, it is easily observable that the ergodic capacity value is very highest for all iterations except the 1st iteration in GA-adaptive mutation and thus proposed work facilitates good ergodic capacity results with the best transmit antennas subset selection for the reduction in hardware complexity by providing the adaptive mutation with best fitness evaluation value for all mutation points.

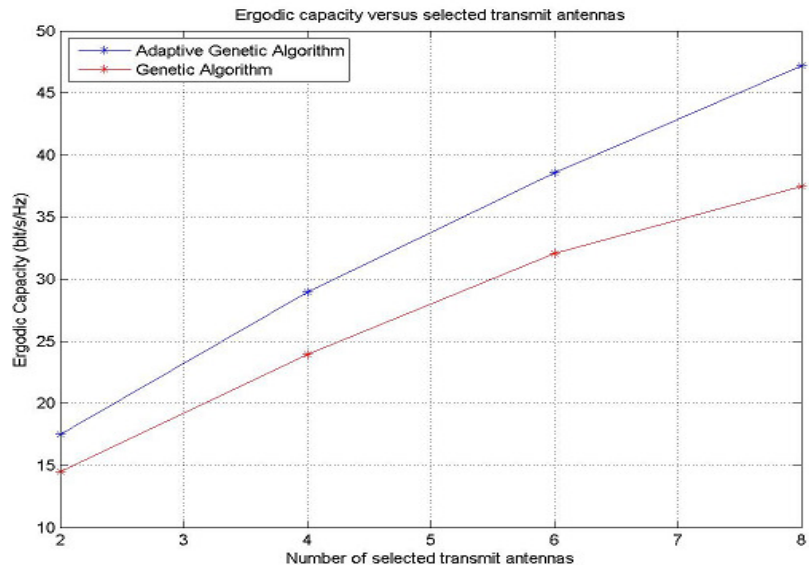

Figure 7 Comparison graph of GA-mutation & GA-adaptive mutation - Ergodic capacity versus number of selected optimal transmit antennas for $N_t = 10, N_r = 10 \ and \ \rho = 15dB$

The last graph in figure 8 displays the comparison performance results of GA in solution convergence while using adaptive mutation and mutation (existing GA). The results of section 4.1, shows the performance results of optimal transmit antenna selection using the proposed GA-adaptive mutation is evaluated with the 10 total number of transmit antennas and 8 array subsets of transmit antennas at the rate $\rho = 15dB$. From these results, the observation is that except the 1$^{st}$ iteration in GA-adaptive mutation, all the other iterations give highest ergodic capacity value. But in the existing GA-mutation work in MIMO-OFDM gives low value of ergodic capacity until it reaches the 10$^{th}$ iteration of GA-mutation. After the 10$^{th}$ iteration, the existing GA-mutation work provides 37.4518 bits/s/ Hz, which is 9.9544 bits/s/Hz lower than the proposed work value (47.4062 bits/s/Hz). On the whole, it is  recognized that the proposed work is best for the effective selection of optimal transmit antennas subsets by providing maximum ergodic capacity and which helps to decrease the complexity in hardware.

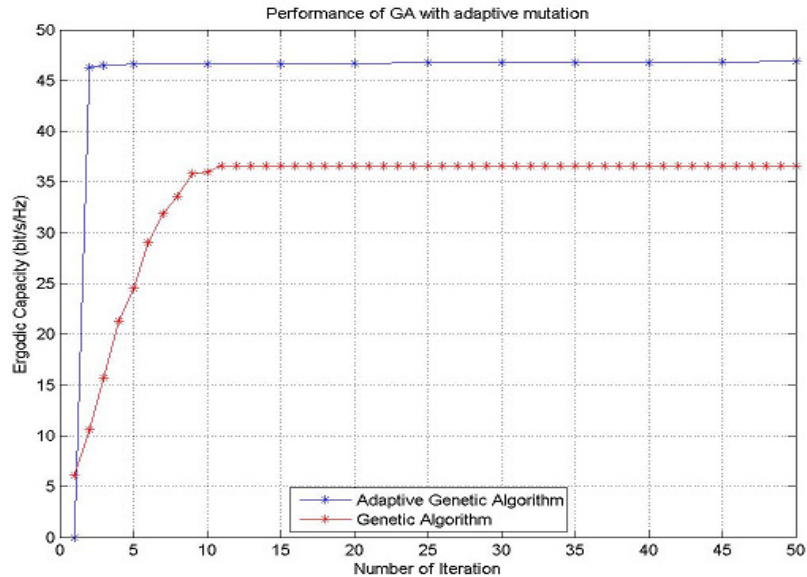

Figure 8 Comparison performance of GA in solution convergence while using adaptive mutation and mutation (existing GA) with  $N_t = 10, N_r = 10 \ and \ \rho = 15dB$

## CONCLUSIONS

In this paper, there is proposal of a straightforward and efficient method using GA for the selections of an optimal subset from an array of transmit antennas. The main aspect of the proposed method is the integration of adaptive approaches into the genetic operator, Mutation. The results of the proposed method have shown that the MIMO-OFDM system attains good capacity by selecting the optimal transmit antenna subset. The simulated results are implemented in the working platform Matlab and the results have demonstrated that their fast convergence to the best possible solution. The proposed work of GA with adaptive mutation facilitates 47.4062 bits/s/Hz of maximum ergodic capacity value. Moreover, the comparison results are also provides that proposed work is better than the existing GA with mutation by providing 9.9544 bits/s/Hz of ergodic capacity value than the existing GA. Thus the vital plan of reducing the hardware complexity with reasonable system capacity in the MIMO-OFDM system has been proficiently achieved by the proposed antenna selection method based on GA with adaptive mutation

## Authors

Nidhi Sindhwani received the B.E. degreein electronics and communication engineering from Mahrishi Dayanand University, Rohtak, India, in 2004 and the M.E. degree in Digital Signal Processing from Mahrishi Dayanand University, Rohtak India, in 2008, she has been pursuing the Ph.D.degree in Wireless Communication at Punjabi university, Patiala, India. Her research focuses on signal processing techniques for communications, including MIMO OFDM, multiuser MIMO communications. Presently she is working as an assistant Professor in ECE/ICE dept.at Amity School of Engineering and Technology, New Delhi.

Dr. Manjit Singh received the Ph.D. degree in Fiber Optics Communication Engineering from PTU Jalandhar, Punjab, India. He has over 39 journal and conference publications,. His research interests span several areas, including Fiber Optics Communication Engineering, Wireless/Mobile Communication.